\documentstyle[twoside,epsf,epsfig]{adacta}
\begin{document}
\def\bEQ{\vspace*{-1.4ex} \begin{equation}}  
\def\eEQ{\vspace*{-1.2ex} \end{equation}} 
\def\bEQA{\vspace*{-1.8ex} \begin{eqnarray}}  
\def\eEQA{\end{eqnarray} \\[-3.25ex]} 
\unitlength1mm
\prvastrana=1
\poslednastrana=6
\def\autor{J. Clausen, M. Dakna, L. Kn\"oll, D.-G. Welsch}
\def\nazov{Conditional quantum state engineering at beam splitter arrays} 

\headings{1}{6}
\title{CONDITIONAL QUANTUM STATE ENGINEERING\\ AT BEAM SPLITTER ARRAYS}
\author{J. Clausen \footnote{\email{clausen@tpi.uni-jena.de}}, M. Dakna, 
  L. Kn\"oll, D.-G. Welsch}
{ Friedrich-Schiller-Universit\"at Jena,\\ Theoretisch-Physikalisches 
Institut,\\ Max-Wien-Platz 1, D-07743 Jena, Germany}

\datumy{30 April 1999}{10 May 1999}
\abstract{
The generation of arbitrary single-mode quantum states from the vacuum by
alternate coherent displacement and photon adding as well as the measurement 
of the overlap of a signal with an arbitrarily chosen quantum state are 
studied. With regard to implementations, the transformation of the quantum 
state of a traveling optical field at an array of beam splitters is considered, 
using conditional measurement. Allowing for arbitrary quantum states of both 
the input reference modes and the output reference modes on which the 
measurements are performed, the setup is described within the concept of 
two-port non-unitary transformation, and the overall non-unitary transformation 
operator is derived. It is shown to be a product of operators, where each 
operator is assigned to one of the beam splitters and can be expressed in terms 
of an $s$-ordered operator product, with $s$ being determined by the beam 
splitter transmittance or reflectance. As an example we discuss the generation 
of and overlap measurement with Schr\"odinger-cat-like states.
}
\section{Introduction}
If two traveling (pulse-shaped) modes of the radiation field are mixed at a 
beam splitter, then the two outgoing modes are in an entangled state in 
general. Therefore, the reduced state of one of them depends on the result of a 
measurement performed on the other. By mixing a signal pulse with a reference 
pulse prepared in a known state and discriminating from all pulses leaving the 
signal output port those corresponding to a particular measurement result in 
the other output port, quantum state engineering can be realized 
\cite{DaknaJacobi,pap2,pap3}.
On the other hand, the overlap of a signal with a chosen quantum state 
may be obtained by mixing the signal mode with a reference mode 
prepared in a quantum state that is 
specific to the overlap and performing a measurement on the outgoing field 
\cite{Barnett}. Hence, novel possibilities of direct quantum state generation 
and measurement are offered, 
provided that the designed reference states can be prepared and the 
required measurements can be realized.

In what follows we present a scheme for the generation of arbitrary quantum 
states of traveling fields, in which coherent and 1-photon Fock states 
are fed into an array of beam splitters and zero-photon measurements are 
performed. We then show how the scheme can be modified in order to measure 
the overlap of an unknown quantum state of a signal mode with an arbitrarily 
chosen quantum state. Whereas in the former case a source for 1-photon Fock 
states should be available, in the latter case only 1-photon Fock state 
detection is required.  

In Section 2 the underlying formalism is outlined and the basic formulas are 
given. The problem of the generation of arbitrary quantum states is considered 
in Section 3, and Section 4 is devoted to the problem of overlap 
measurements. In order to 
give an example, we consider in Section 5 the generation of and measurement of 
overlap with Schr\"odinger-cat-like states. A summary and some concluding 
remarks are given in Section 6.
\section{Conditional quantum state transformation}
Let us consider the state transformation at a beam splitter array. 
As outlined in Fig.1,
\begin{figure}[tbh]
\vspace{2.6ex}
\begin{minipage}[b]{0.55\linewidth}
{\centering\epsfig{figure=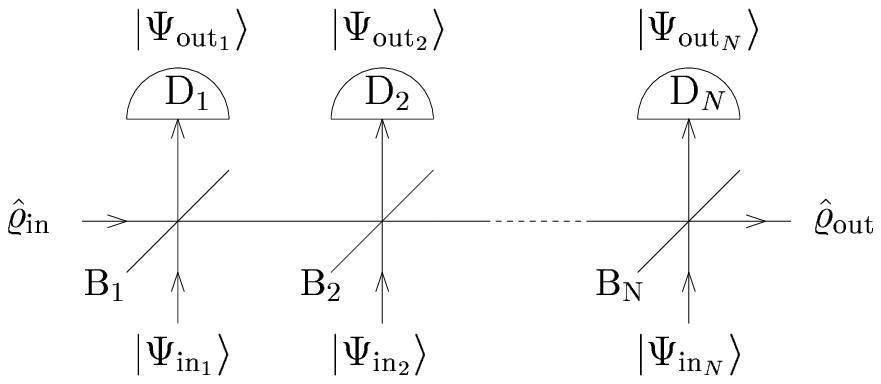,width=0.9\linewidth}}
\end{minipage}\hfill
\begin{minipage}{0.45\linewidth}
{\vspace{-18ex}\footnotesize 
 Fig.~1. Conditional quantum state manipulation at an array of beam splitters 
${\rm B}_k$. The incoming signal state $\hat{\varrho}_{{\rm in}}$ is 
combined with states $|\Psi_{{\rm in}_k}\rangle$, and  
$\hat{\varrho}_{{\rm out}}$ is the signal state generated
under the condition that the outgoing reference modes 
have been measured to be in states $|\Psi_{{\rm out}_k}\rangle$ by means of 
the measuring devices ${\rm D}_k$.
}
\end{minipage}
\vspace{-1.8ex}
\end{figure}
the incoming signal prepared in a state $\hat{\varrho}_{{\rm in}}$ passes 
an array of $N$ beam splitters \mbox{${\rm B}_1,$ $\!\ldots,$ $\!{\rm B}_N$} 
at which it is mixed with reference input modes in states 
\mbox{$|\Psi_{{\rm in}_1}\rangle,$ $\!\ldots,$ $\!|\Psi_{{\rm in}_N}\rangle$}. 
When the measuring devices ${\rm D}_1,$ $\!\ldots,$ $\!{\rm D}_N$ detect the 
reference output modes in states \mbox{$|\Psi_{{\rm out}_1}\rangle,$ 
$\!\ldots,$ $\!|\Psi_{{\rm out}_N}\rangle$}, 
then the conditional signal output state reads 
\bEQ
  \hat{\varrho}_{{\rm out}}=\frac{1}{p}\,\hat{Y}
  \hat{\varrho}_{{\rm in}}
  \hat{Y}^\dagger,
\eEQ
where
\bEQ
  p={\rm Tr}\!\left(\hat{Y}\hat{\varrho}_{{\rm in}}\hat{Y}^\dagger\right)
\eEQ
is the probability of generating $\hat{\varrho}_{{\rm out}}$. 

The non-unitary transformation operator 
\bEQ
  \hat{Y}=\hat{Y}_N\cdots\hat{Y}_2\hat{Y}_1
\eEQ
is the product of the individual conditional operators
\bEQ
  \hat{Y}_k=\langle\Psi_{{\rm out}_k}|\hat{U}_k|
  \Psi_{{\rm in}_k}\rangle
  \,,
  \label{Yk}
\eEQ
where 
\bEQ
  \hat{U}_k = e^{{\rm i}(\varphi_T + \varphi_R)\hat{L}_z}\,
  e^{2{\rm i}\vartheta\hat{L}_y}\, e^{{\rm i}(\varphi_T 
  -\varphi_R)\hat{L}_z} = T^{\hat{n}}\,e^{-R^*\hat{a}^\dagger_k
  \hat{a}}\,e^{R\hat{a}^\dagger\hat{a}_k}T^{-\hat{n}_k}
\eEQ
is the unitary transformation operator of the beam splitter 
${\rm B}_k$ in Fig.~1, with
$\hat{L}_y$ $\!=$ $\!{\rm i}(\hat{a}_k^\dagger\hat{a}$ $\!-$ 
$\!\hat{a}^\dagger\hat{a}_k)/2$ 
and $\hat{L}_z$ $\!=$ $\!(\hat{n}$ $\!-$ $\!\hat{n}_k)/2$ 
\cite{Campos,Daknaadded}. Here,
$T$ $\!=$ $\!\cos\vartheta e^{{\rm i}\varphi_T}$ and 
$R$ $\!=$ $\!\sin\vartheta e^{{\rm i}\varphi_R}$ are the 
transmittance and reflectance, respectively. 
The operators $\hat{Y}_k$ can be 
represented as $s$-ordered operator products \cite{pap3}
\bEQ
  \hat{Y}_k = \left\{\hat{F}(R\hat{a}^\dagger)\,
  \hat{G}^\dagger\!\left(-\frac{R}{T^*}\hat{a}^\dagger\right)
  \right\}_sT^{\hat{n}},
  \label{order}
\eEQ
where the operators $\hat{F}$ and $\hat{G}$, respectively,
generate $|\Psi_{{\rm in}_k}\rangle$ and 
$|\Psi_{{\rm out}_k}\rangle$ from the vacuum, 
\bEQA 
  |\Psi_{{\rm in}_k}\rangle=\hat{F}(\hat{a}_k^\dagger)|0\rangle_k\,,
  \quad
  |\Psi_{{\rm out}_k}\rangle=\hat{G}(\hat{a}_k^\dagger)|0\rangle_k\,,
\eEQA 
and the ordering parameter $s$ is determined by the absolute value of the beam 
splitter reflectance as
\bEQ
  s=\frac{2}{|R|^2}-1 .
\eEQ
Note that the ordering procedure in (\ref{order}) can be omitted if 
$|\Psi_{{\rm in}_k}\rangle$ or $|\Psi_{{\rm out}_k}\rangle$ is a coherent 
state \cite{pap3}, since for
\bEQA 
  |\Psi_{{\rm in}_k}\rangle=\hat{D}_k(\alpha)\hat{F}(\hat{a}_k^\dagger)
  |0\rangle_k\,,
  \quad
  |\Psi_{{\rm out}_k}\rangle=\hat{D}_k(\beta)\hat{G}(\hat{a}_k^\dagger)
  |0\rangle_k\,
\eEQA 
we have
\bEQ
  \hat{Y}_k = \hat{D}\!\left(\frac{\alpha-T\beta}{R^*}\right)\,
  \hat{Y}_k\!\left(^{\alpha=0}_{\beta=0}\right)\,
  \hat{D}\!\left(\frac{\beta-T^*\alpha}{R^*}\right).
  \label{noorder}
\eEQ
\section{Generation of truncated quantum states $|\Psi\rangle$}
Each quantum state $|\Psi\rangle$ 
that is composed of a finite 
number of Fock states $|n\rangle$ can be written as 
\bEQ
  |\Psi\rangle=\sum_{n=0}^N\,\psi_n|n\rangle
  =\frac{\psi_N}{\sqrt{N!}}\prod_{k=1}^N
  (\hat{a}^\dagger-\beta_k^*)|0\rangle,
  \label{state}
\eEQ
where $\beta_1,$ $\!\ldots,$ $\!\beta_N$ denote the $N$ solutions of the 
equation $\langle\Psi|\beta\rangle\equiv\langle\Psi|\hat{D}(\beta)|0\rangle=0$. 
This reveals that $|\Psi\rangle$ can be generated from the vacuum state by 
alternate displacement and photon adding, as outlined in Fig.~2 \cite{pap2}. 
\begin{figure}[tbh]
\vspace{1ex}
\begin{minipage}[b]{0.74\linewidth}
{\centering\epsfig{figure=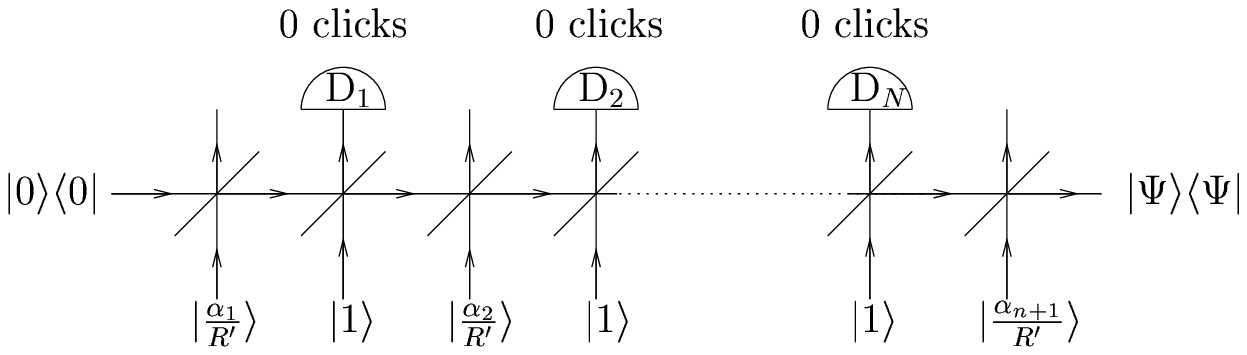,width=0.9\linewidth}}
\end{minipage}\hfill
\begin{minipage}{0.26\linewidth}
{\vspace{-15ex}\footnotesize 
 Fig.~2. Generation of trun\-cated states $|\Psi\rangle$ from the vacuum by 
alternate displacement and photon adding. The coherent displacements are 
achieved by combination 
\hfill 
with 
\hfill 
highly%
}
\end{minipage}\\[1ex]
\noindent
{\footnotesize 
excited coherent states 
$|\alpha/R^\prime\rangle$ at highly transmitting beam splitters
($T^\prime\rightarrow1$), and photon adding is realized by combination with 
one-photon Fock states $|1\rangle$ and measuring zero photons in the outgoing 
reference modes with photodetectors D$_k$.
}
\vspace{-1ex}
\end{figure}
A coherent displacement $\hat{D}(\alpha)$ can be realized by mixing the mode 
with a reference mode in a strong coherent state $|\alpha/R^\prime\rangle$ at a 
highly transmitting beam splitter \mbox{($T^\prime$ $\!\rightarrow$ $\!1$)} 
\cite{Paris}, and photon adding is achieved by mixing the mode with a reference
mode in a Fock state $|1\rangle$ and measuring zero photons in the output 
detection channel (detectors ${\rm D}_k$ in Fig.~2) \cite{Daknaadded}. 

We assume that all the beam splitters used for photon adding have the same 
transmittance $T$ and reflectance $R$. The non-unitary transformation operator 
then reads as 
\bEQ
  \hat{Y}=\hat{D}(\alpha_{N+1})\hat{Y}_N\hat{D}(\alpha_N)\cdots
  \hat{Y}_1\hat{D}(\alpha_1)\,,
\eEQ
where
\bEQ
  \hat{Y}_k=R\hat{a}^\dagger T^{\hat{n}}\,,
  \label{Ykgen}
\eEQ
and the complex parameters $\alpha_1$,\ldots, $\alpha_{N+1}$ are determined 
from the equation
\bEQ
  \frac{\hat{Y}|0\rangle\langle0|\hat{Y}^\dagger}{\|\hat{Y}|0\rangle\|^2}=
  |\Psi\rangle\langle\Psi|
\eEQ
as $\alpha_k$ $\!=$ $\!T^{*N+1-k}(\beta_{k-1}$ $\!-$ $\!\beta_k)$ for  
$k$ $\!=$ $\!2,\ldots,N+1$
($\beta_{N+1}$ $\!=$ $\!0$), and $\alpha_1$ $\!=$ $\!-\sum_{l=1}^NT^{-l}
\alpha_{l+1}$. 
The probability of generating the desired state 
$\hat{\varrho}_{{\rm out}}=|\Psi\rangle\langle\Psi|$,
i.e. the probability that all $N$ detectors register zero photons, 
is given by 
\bEQ
  \|\hat{Y}|0\rangle\|^2=\frac{N!}{|\psi_N|^2}
  \frac{|R|^{2N}}{|T|^{N(1-N)}}\, \exp\!\left[-|R|^2\sum_{k=1}^N
 \bigg |\frac{\sum_{l=1}^k|T|^{2l}(\beta_{N+2-l}\!-\!\beta_{N+1-l})}
  {T^{k+2}}\bigg|^2\right]
  \label{prob}
\eEQ
and decreases rapidly with increasing $N$. Nevertheless, for small $N$ this 
scheme offers a way to generate specific traveling quantum states, given the 
possibility to prepare 1-photon Fock states.
\section{Measuring arbitrary overlaps}
One fundamental task in quantum mechanics is to find the probability that 
a specific quantum state $|\Psi\rangle$ is contained in a state 
$\hat{\varrho}_{{\rm in}}$ of a given system. With regard to traveling waves, 
this overlap $\langle\Psi|\hat{\varrho}_{{\rm in}}|\Psi\rangle$ 
can be measured for a given state $|\Psi\rangle$ of the type (\ref{state})
as outlined in Fig.~3 \cite{pap4}. 
\begin{figure}[tbh]
\vspace*{2.5ex}
\begin{minipage}[b]{0.55\linewidth}
{\centering\epsfig{figure=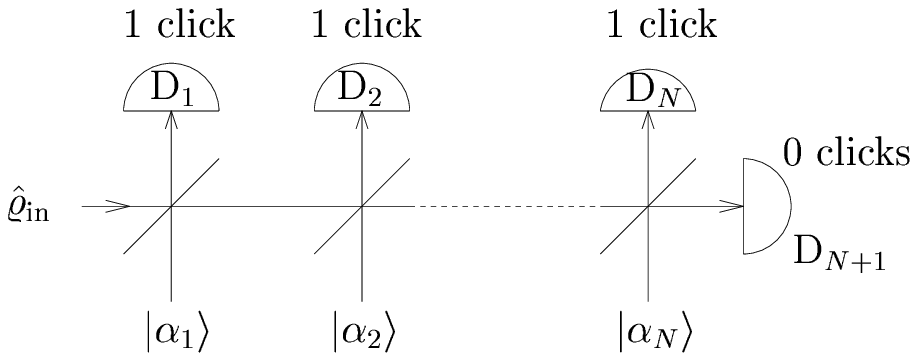,width=0.9\linewidth}}
\end{minipage}\hfill
\begin{minipage}{0.45\linewidth}
{\vspace*{-15ex}
\footnotesize 
 Fig.~3. Measurement of the overlap 
$\langle\Psi|\hat{\varrho}_{{\rm in}}|\Psi\rangle$ of a signal state 
$\hat{\varrho}_{{\rm in}}$ with a desired state $|\Psi\rangle$ by successive
combination with appropriately chosen coherent states $|\alpha_1\rangle,$ 
$\!\ldots,$ $\!|\alpha_N\rangle$ at beam splitters and 
measurement of the relative frequency of
detecting simultaneously $1$ photon at the photodetectors 
${\rm D}_1,$ $\!\ldots,$ ${\rm D}_N$ and 0 photons at 
${\rm D}_{N+1}$.
}
\end{minipage}
\vspace{-0.5ex}
\end{figure}
The signal mode in state $\hat{\varrho}_{{\rm in}}$ is mixed with
reference modes in coherent states $|\alpha_1\rangle,$ $\!\ldots,$ 
$\!|\alpha_N\rangle$ at an array of $N$ beam splitters. Photodetectors 
${\rm D}_1,$ $\!\ldots,$ $\!{\rm D}_{N+1}$ perform 
photon number measurements at the output modes. 
The joint probability $p(1,1;2,1;\ldots;N,1;N+1,0)$ that 
each of the detectors
\mbox{${\rm D}_1,$ $\!\ldots,$ $\!{\rm D}_{N}$} 
registers one photon and ${\rm D}_{N+1}$ none is then given by 
\bEQ
  p(1,1;2,1;\ldots;\;N,1;\;N+1,0)=
  \langle0|\hat{Y}\hat{\varrho}_{{\rm in}}\hat{Y}^\dagger|0\rangle
  \label{joint}
\eEQ
with
\bEQ
  \hat{Y}=\hat{Y}_N\cdots\hat{Y}_2\hat{Y}_1\,,
\eEQ
where for identical beam splitters
\bEQ
  \hat{Y}_k=-R^*\hat{D}\left(\frac{\alpha_k}{R^*}\right)T^{\hat{n}}
  \hat{a}\hat{D}\left(-\frac{T^*}{R^*}\alpha_k\right),
  \label{Ykpro}
\eEQ
see (\ref{Yk}) as well as (\ref{order}) and (\ref{noorder}). This 
expression reveals that the signal is manipulated by alternate displacement 
and photon subtraction. If we now choose the arguments 
$\alpha_k$ ($k$ $\!=$ $\!1,\ldots,N$) of the coherent states as
$\alpha_k$ $\!=$ $\!(R^*/T^{*k})\sum_{l=1}^k|T|^{2l-1}(\beta_l$ $\!-$ 
$\!\beta_{l-1})$, with \mbox{$\beta_0$ $\!=$ $\!0$}, so that for a chosen 
$|\Psi\rangle$ the relation 
\bEQ
  \frac{\hat{Y}^\dagger|0\rangle\langle0|\hat{Y}}
  {\|\hat{Y}^\dagger|0\rangle\|^2}=|\Psi\rangle\langle\Psi|
\eEQ
is valid, then from (\ref{joint}) it is seen that the sought
overlap can be obtained from the joint probability as
\bEQ
  \langle\Psi|\hat{\varrho}_{{\rm in}}|\Psi\rangle=
  \frac{p(1,1;2,1;\ldots;N,1;N+1,0)}
  {\|\hat{Y}^\dagger|0\rangle\|^2}\,.
\eEQ
The denominator $\|\hat{Y}^\dagger|0\rangle\|^2$ is
given by (\ref{prob}), with $(\beta_{N+2-l}$ $\!-$ $\!\beta_{N+1-l})$ 
being replaced with $(\beta_l$ $\!-$ $\!\beta_{l-1})$. It may be
regarded as being the ``fidelity'' of the measurement, 
since it is a measure of the maximal occurence of the detection 
coincidences. Obviously, it is equal to 
$p(1,1;2,1;\ldots;N,1;N+1,0)$ in the particular case when signal 
and measured state coincide, 
$\hat{\varrho}_{{\rm in}}$ $\!=$ $\!|\Psi\rangle\langle\Psi|$.
\section{Schr\"odinger-cat-like states}
The schemes in Figs.~2 and 3 can be simplified if some of the
$\beta_k$ in (\ref{state}) are equal:
\bEQ
  |\Psi\rangle=\frac{\psi_N}{\sqrt{N!}}\prod_{l=1}^M
  (\hat{a}^\dagger-\beta_l^*)^{d_l}|0\rangle
\eEQ
with $M$ $\!<$ $\!N$. In this case it is possible to add $d_k$ photons 
at once in Fig.~2 by using $M$ detectors and combining with 
 Fock states $|d_k\rangle$ or 
to subtract $d_k$ photons at each beam splitter in Fig.~3 by using 
$M$ $\!+$ $\!1$ detectors and measuring the relative frequency of the 
event $(1,d_1;\ldots;M,d_M;M+1,0)$. All calculations are analogous if 
$R\hat{a}^\dagger$ in (\ref{Ykgen}) is replaced with 
$(R\hat{a}^\dagger)^{d_k}/\sqrt{d_k!}$ and $-R^*\hat{a}$ in (\ref{Ykpro}) 
is replaced with $(-R^*\hat{a})^{d_k}/\sqrt{d_k!}$.

As an example let us consider the states
\bEQ
  |\Psi_n^{\alpha,\beta}\rangle=\frac{1}{\sqrt{{\cal N}}}\,\hat{D}(\gamma_3)
  \hat{a}^{\dagger n}\hat{D}(\gamma_2)\hat{a}^{\dagger n}\hat{D}(\gamma_1)
  |0\rangle,
\eEQ
where $\gamma_1$ $\!=$ $\!{\rm i}(\beta$ $\!-$ $\!\alpha)/2$, 
$\gamma_2$ $\!=$ $\!{\rm i}(\alpha-\beta)$, 
$\gamma_3$ $\!=$ $\![(1-{\rm i})\alpha$ $\!+$ $\!(1+{\rm i})\beta]/2$, 
and the normalization factor is ${\cal N}$ $\!=$ $\!(4^nn!/\sqrt{\pi})$
$\!\mbox{$\Gamma(n+1/2)$}$ $\!_1F_2[-n,1/2-n,1,|\alpha-\beta|^4/64]$. From 
the above considerations it is clear that these states can be generated 
using 2 detectors, and the overlap of a signal state with such states can be 
measured using 3 detectors. The states $|\Psi_n^{\alpha,\beta}\rangle$ reveal 
the interesting property that for increasing $n=|\alpha-\beta|^2/4$ they
approach superpositions of coherent states \cite{pap4}, 
$|\Psi_\infty^{\alpha,\beta}\rangle\langle\Psi_\infty^{\alpha,\beta}|=
|\Psi^{\alpha,\beta}\rangle\langle\Psi^{\alpha,\beta}|$ with
\bEQ
  |\Psi^{\alpha,\beta}\rangle=\frac{1}{\sqrt{2}}\left(
  |\alpha\rangle+|\beta\rangle \right)
\eEQ
(note that 
$|\langle\Psi_{n=3}^{\alpha,\beta}|\Psi^{\alpha,\beta}\rangle|^2$
$\!>$ $\!0.95$).
This offers the possibility of generating Schr\"odinger-cat-like states,
provided that two $n$-photon Fock states are available. On the other hand, 
measuring overlaps with Schr\"odinger-cat-like states allows one,
in principle, to reconstruct the signal state \cite{Freyberger}. 

It is worth noting that choosing a squeezed coherent signal state 
offers the possibility of measuring the field strength statistics 
of a Schr\"odinger cat without need to generate the cat state.
Note that squeezed coherent states approach field strength states for 
sufficiently strong squeezing.

 Finally, it should be pointed out that the probability of generation 
of the states $|\Psi_n^{\alpha,\beta}\rangle$ and the fidelity of 
measurement of the overlap with them are equal. For large $n$, they
decrease exponentially with increasing $n$:
\bEQ
  p=\frac{|2R^2T|^{2n}}{n\pi}\,\exp\!\left[n\left(1\!-\!
  \left|\frac{R}{T}\right|^2(1\!+\!|T|^{-2}(1-2|T|^2)^2)\right)\right].
  \label{p}
\eEQ
\section{Conclusion}
We have discussed conditional quantum state engineering at beam splitter 
arrays. We have presented a scheme for the generation of arbitrary quantum 
states which requires coherent states and 1-photon Fock states
and zero-photon detection. 
 Further, we have given a scheme for the measurement of
the overlap of an unknown signal state with an arbitrary quantum state
which requires coherent states and $0$- and $1$-photon detections.
 For the two schemes we have calculated the probabilities of state 
generation and overlap measurement. Finally, we have shown how the 
schemes can be simplified under special conditions. As an example we 
have considered the generation of (and overlap measurement with) states 
which approach superpositions of two arbitrary coherent states.

\medskip
\noindent {\bf Acknowledgements}\\[1ex]
This work was supported by the Deutsche Forschungsgemeinschaft.

\end{document}